# Premier décryptage du règlement européen sur l'intelligence artificielle (AI Act) : Vers un standard mondial de l'IA de confiance ?

Marion Ho-Dac, Professeure de droit privé à l'Université d'Artois,
Centre Droit Éthique et Procédures (CDEP) UR2471

**Résumé** : *Le règlement sur l'intelligence artificielle (AI Act) est entré en vigueur le 1er août 2024 dans l'Union européenne. Il s'agit d'un texte-clé tant pour les citoyens placés au cœur des technologies d'IA que pour l'industrie active dans le marché intérieur. L'AI Act impose une mise en conformité progressive pour les organisations – privées comme publiques –, intervenant dans la chaine de valeur mondiale des systèmes et modèles d'IA mis sur le marché et utilisés dans l'Union. Alors que le texte est inédit à l'échelle internationale de par son emprise règlementaire horizontale et contraignante, sa force d'attraction mondiale au soutien de l'IA de confiance est l'un de ces enjeux majeurs.*

L'Union européenne (ci-après « l'Union ») vient d'adopter un cadre juridique uniforme applicable aux systèmes d'intelligence artificielle (ci-après « systèmes d'IA ») dans le marché intérieur[1]. Le règlement (UE) 2024/1689 dit « *règlement sur l'intelligence artificielle* » (ci-après « AI Act ») fera date car il constitue le premier texte de nature horizontale fixant des exigences règlementaires contraignantes pour les principaux opérateurs[2] – y compris les autorités publiques[3] – de la chaîne de valeur mondiale des systèmes d'IA les plus à risque mis sur le marché ou utilisés dans l'Union[4]. D'un côté, il s'inscrit dans l'écosystème dynamique de la gouvernance mondiale des technologies d'IA[5] et en propose une traduction législative fondée sur la co-régulation[6]. De l'autre, il impose sa marque règlementaire en prenant appui sur la politique européenne de sécurité des produits et la méthodologie du nouveau cadre législatif (« *New Legislative Framework* », ci-après « NLF ») fondée sur l'approche par les risques[7]. Il met en place, à titre principal, un schéma de conformité *ex ante* alimenté par des normes techniques et un cadre *ex post* de surveillance du marché, tout au long du cycle de vie des systèmes d'IA. Le concept de « risque » adopté par l'AI Act s'émancipe toutefois partiellement, dans sa lettre, de l'acquis NLF[8] : il inclut, aux côtés des risques pour « *la santé et la sécurité* », ceux visant « *les droits fondamentaux* » et, plus largement encore, « *les valeurs de l'Union* » à l'instar du respect de la démocratie et de l'État de droit[9]. Cela marque une inclinaison législative « humaniste » dont la portée devra être finement scrutée (et certainement défendue) au stade de la mise en œuvre du règlement[10].

---

[1] Règlement (UE) 2024/1689 du Parlement européen et du Conseil du 13 juin 2024 établissant des règles harmonisées concernant l'intelligence artificielle et modifiant les règlements (CE) n° 300/2008, (UE) n° 167/2013, (UE) n° 168/2013, (UE) 2018/858, (UE) 2018/1139 et (UE) 2019/2144 et les directives 2014/90/UE, (UE) 2016/797 et (UE) 2020/1828, JO L, 2024/1689, 12 juillet 2024, ci-après « AI Act ».
[2] Selon l'article 3, point 8, de l'AI Act, l'opérateur désigne tout à la fois le fournisseur, le fabricant de produits, le déployeur, le mandataire (du fournisseur), l'importateur ou le distributeur. Au sein de l'AI Act, les deux protagonistes centraux sont, d'un côté, le fournisseur, entendu comme celui qui développe (ou fait développer) le système et le commercialise et, de l'autre, le déployeur qui l'utilise à titre professionnel.
[3] Cette appréhension large des « opérateurs », personnes physiques ou morales, y compris des autorités publiques, des agences ou d'autres organismes (y compris de l'Union), découle des définitions de « fournisseur » et de « déployeur » (qui sont des professionnels au sens du droit du marché), pts 3 et 4, art. 3, AI Act.
[4] Sur ce texte final, v. not. J. Sénéchal, « L'AI Act dans sa version finale – provisoire –, une hydre à trois têtes », *Dalloz actualité*, 11 mars 2024 ; N. A. Smuha & K. Yeung, "The European Union's AI Act: beyond motherhood and apple pie?", 24 Juin 2024, disponible sur SSRN ; O. Tambou, Règlement européen sur l'IA: un texte voté mais encore des ambiguïtés, *The Conversation*, 24 mars 2024
[5] V. M. Chinen, *The International Governance of Artificial Intelligence*, 2023, Cheltenham UK/Northampton MA USA, Edward Elgar Publishing.
[6] Au sens de la combinaison règlementaire, prévue par la loi, entre règles législatives et normes d'entreprises.
[7] V. « Guide bleu » relatif à la mise en œuvre de la règlementation de l'UE sur les produits 2022, 2022/C 247/01, JO C 247 du 29 juin 2022 et cons. 26, AI Act.

L'AI Act est donc un texte-clé pour les citoyens qui sont placés au cœur des technologies d'IA dans leur dimension innovante mais aussi à risque pour les droits et libertés. Il est également incontournable pour l'industrie, l'Union étant l'un des principaux marchés mondiaux de consommation. Il va s'agir pour les organisations – privées comme publiques –, dans les mois et années à venir, de se mettre en conformité de quelques 113 articles, 180 considérants et 13 annexes[11], auxquels viendront s'ajouter de nombreux documents juridiques et techniques que la Commission européenne et d'autres organes de l'écosystème de l'AI Act sont appelés à produire.

Le décryptage du règlement implique l'étude successive de son champ d'application (I), des régimes de conformité qu'il instaure (II) et de la gouvernance multi-niveaux qu'il met en place (III).

## I) Le champ d'application de l'AI Act

L'AI Act est doté d'un vaste champ d'application. D'une part, il retient une définition souple des systèmes d'IA inspirée de celle élaborée dans l'enceinte internationale de l'OCDE[12]. Il s'agit de tout « *système automatisé […] conçu pour fonctionner à différents niveaux d'autonomie et [pouvant] faire preuve d'une capacité d'adaptation après son déploiement, et qui, pour des objectifs explicites ou implicites, déduit, à partir des entrées qu'il reçoit, la manière de générer des sorties telles que des prédictions, du contenu, des recommandations ou des décisions qui peuvent influencer les environnements physiques ou virtuels* »[13]. D'autre part, il affirme sa dimension horizontale au sein du marché européen des produits, associée au principe de *lex specialia* du fait de la particularité de son objet – les technologies avancées d'IA –[14]. A l'examen cependant, quelques nuances s'imposent. Tant la taxonomie des systèmes et modèles d'IA élaborée par l'AI Act (A) que le marché qu'il entend couvrir (B) connaissent des limitations et des exceptions qui atténuent la portée englobante, à vocation mondiale, du texte.

---

[8] V. la notion de produit présentant un risque dans le règlement (UE) 2019/1020 du 20 juin 2019 sur la surveillance du marché et la conformité des produits, JO L 169, 25 juin 2019, spéc. art. 3, pt 19.
[9] Cons. 1 et 2, et art. 1, §1, AI Act.
[10] V. la référence à la « Déclaration européenne sur les droits et principes numériques pour la décennie numérique » qui devrait être « *prise en compte* » par les règles de l'AI Act, cons. 7.
[11] Le règlement est entré en vigueur le 1er août 2024. Il sera applicable, selon l'article 113, entre le 2 février 2025 et le 2 août 2027, selon les dispositions du texte. Par défaut, il le sera le 2 août 2026.
[12] https://oecd.ai/fr/ai-principles
[13] Art. 3, pt 1, AI Act.
[14] Cons. 46 et 64, AI Act.

**A) La taxonomie des systèmes et modèles d'IA**

La taxonomie des systèmes d'IA élaborée par l'AI Act est complexe dans sa délimitation fine[15]. A côté des pratiques interdites car présentant des risques inacceptables (1) et des systèmes d'IA à haut risque (2), une catégorie *ad hoc* pour les modèles d'IA à usage général (ci-après « modèles d'IAUG ») (3) a été insérée en fin de négociations, sous la pression de la diffusion fulgurante de l'IA générative auprès du grand public. Il n'existe en revanche aucune trace lisible d'une catégorie de systèmes d'IA à risque modéré qui figure généralement dans la pyramide des risques associée à l'AI Act[16].

**1) Les pratiques d'IA interdites**

L'article 5, §1 de l'AI Act prévoit une liste de huit cas d'usage donnant lieu à une interdiction de plein droit, à l'instar des systèmes d'IA capables de sonder les émotions ou d'influencer le consentement humain, et des pratiques de *scoring* social, de prédiction en matière d'infraction pénale et de biométrie. Ces interdictions sont fondées sur la contrariété de ces pratiques aux valeurs de l'Union à commencer par le respect de la dignité humaine et de la liberté[17]. L'Union donne ainsi à voir au reste du monde sa vision politique de la vie en société. Cependant, cette démonstration « quasi-constitutionnelle » est mise à mal. Ces interdictions sont toutes, et parfois très largement, conditionnées et souvent amputées par des exceptions. Elles ne sont donc pas aisément prévisibles alors même que leur qualification est laissée à l'appréciation des fournisseurs. Ainsi, par exemple, les États membres ont réussi à maintenir une forte exception en matière répressive (« *law enforcement* ») afin de permettre l'utilisation de systèmes biométriques qui peuvent pourtant porter gravement atteinte à l'intégrité humaine dans nombre de situations[18].

**2) Les systèmes d'IA à haut risque**

L'article 6 de l'AI Act distingue deux catégories de systèmes d'IA à haut risque entendus limitativement comme ceux « […] *qui ont une incidence préjudiciable substantielle sur la santé, la sécurité et les droits fondamentaux des citoyens dans l'Union* »[19]. La première découle de l'article 6, §1, lu en combinaison avec l'annexe I qui prévoit une double liste (A et B) de législations d'harmonisation de l'Union, la liste A étant directement issue du NLF. Pour être qualifié de système à haut risque, il faut que le produit ou le composant de sécurité d'un produit intégrant l'IA soit régi par l'une des législations d'harmonisation de cette annexe I et soit soumis, selon cette législation, à un « module » de conformité exigeant l'évaluation par un tiers, c'est-à-dire par un organisme d'évaluation de conformité.

---

[15] A ce titre, l'AI Act confère un rôle majeur à la Commission européenne qui sera en charge d'élaborer des lignes directrices pour clarifier cette taxonomie et, en vue de son adaptation aux évolutions technologiques rapides, d'adopter des actes délégués. V. art. 96, §1 et 97, AI Act.
[16] Pour une reconstruction du visuel de la pyramide à l'aune de la version finale de l'AI Act : Th. Christakis & Th Karathanasis, "Tools for navigating the EU AI Act: (2) Visualisation pyramid", 7 Mars 2024 (en ligne).
[17] Au sens de l'article 2 TUE. V. cons. 28, AI Act.
[18] Art. 5, §2 à 7, AI Act.
[19] Cons. 46, AI Act.

La seconde catégorie de système à haut risque prend appui sur l'article 6, §2 de l'AI Act, lu en combinaison avec l'annexe III qui prévoit une liste de huit domaines, avec 25 cas d'usage présentant des niveaux de granularité plus ou moins avancés, en matière de biométrie, d'infrastructures critiques, de justice répressive, d'éducation et d'emploi, de services essentiels, de migration et d'administration de la justice[20]. La catégorie est complexe car elle embrasse une grande diversité d'usage de l'IA, ayant chacun leur spécificité, parfois en prolongement des pratiques interdites (reconnaissance des émotions ou biométrie en particulier) et appelant donc à une articulation avec celles-ci[21].

Dans le même temps, l'article 6, §3 de l'AI Act prévoit une certaine flexibilité en faveur des fournisseurs quant à la qualification de système d'IA à haut risque. Il introduit une sorte de règle *de minimis* permettant au fournisseur d'exclure un système donné de la catégorie « haut risque » si ledit système ne présente pas de risque significatif. Il s'agit certes d'alléger la charge règlementaire des fournisseurs ; pour autant, fallait-il leur laisser une telle latitude ?

### 3) Les modèles d'IAUG

En tant que « nouveaux venus » dans le règlement, les modèles d'IAUG constituent une catégorie *ad hoc* soumise à un régime propre, renforcé en cas de « *risques systémiques* » associés aux modèles les plus puissants[22]. L'AI Act opte pour une définition fonctionnelle de ces modèles, à l'aune de « *[leur] généralité et [de leur] capacité [à] exécuter de manière compétente un large éventail de tâches distinctes* »[23]. Ils sont généralement commercialisés au moyen de bibliothèques, d'interfaces de programmation d'applications (API) ou de téléchargements ; ils sont modifiables et ont vocation à être intégrés dans un système d'IA[24]. Ce sont les grands modèles d'IA générative venus d'outre-Atlantique – à l'instar de « *GPT-3* » puis « *-4* » de l'entreprise *OpenAI*, intégré par le célèbre agent conversationnel *Chat-GPT* – qui constituent l'arrière-fond emblématique de cette catégorie. Ces modèles peuvent présenter des risques systémiques, c'est-à-dire susceptibles d'avoir une incidence significative et à grande échelle dans l'Union, « *sur la société dans son ensemble* », au-delà d'une atteinte aux seuls droits individuels[25].

Le critère-clé de qualification d'un modèle d'IAUG présentant des risques systémiques est celui de ses « *capacités d'impact élevées* »[26]. Ce critère, de nature technique, s'analyse à partir d'éléments méthodologiques et de référence listés à l'annexe XIII de l'AI Act ; il est présumé réalisé lorsque le volume de calcul utilisé pour l'apprentissage du modèle dépasse un certain seuil[27]. La Commission européenne pourra d'office opérer cette qualification[28], ce qui démontre la sensibilité de cette catégorie ; en principe, il reviendra aux fournisseurs de ces modèles de les notifier à la Commission. Afin de ne pas freiner l'innovation, le règlement prévoit que les fournisseurs pourront également demander une « disqualification » en démontrant par « *des arguments suffisamment étayés* » que leur modèle n'est pas porteur de risques systémiques[29].

---

[20] L'AI Act procède de la sorte à une forme d'européanisation indirecte des droits nationaux dans des domaines sensibles comme la justice, l'éducation ou la matière électorale, à partir du moment où le recours aux technologies d'IA pénètrent ces secteurs.
[21] La Commission sera chargée d'élaborer (art. 96) des lignes directrices pour clarifier à la fois le champ d'application et le régime de cette catégorie (art. 6, §5).
[22] Art. 51 et art. 3, pt 65, AI Act.
[23] Cons. 97, AI Act.
[24] *Ibid*.
[25] Définis à l'art. 3, pt 66, AI Act.
[26] Art. 51, §1, a), AI Act.

**B) Le marché des systèmes et modèles d'IA**

La délimitation matérielle (1) et spatiale (2) du marché que l'AI Act entend réguler est révélatrice de la portée mondiale que l'Union ambitionne pour ce texte. Cela ressort distinctement des volets règlementaires de conformité et de *public enforcement* du règlement, à l'exclusion de tout schéma transnational de *private enforcement* confié à d'autres instruments législatifs[30].

**1) Les contours matériels de l'AI Act**

Tout d'abord, l'AI Act se compose de trois grands corpus normatifs[31]. Le premier regroupe des « *règles harmonisées concernant la mise sur le marché, la mise en service et l'utilisation de systèmes d'IA dans l'Union* »[32] et inclut, à un niveau de granularité supérieur, des dispositions interdisant certaines pratiques d'IA ainsi que des exigences spécifiques pour les systèmes à haut risque et les modèles d'IA. Le second corpus est constitué d'un schéma assez dense de *public enforcement* comprenant, d'une part, les règles de surveillance de marché propres au marché européen des produits, renforcées, d'autre part, par un cadre de gouvernance « fédérale » inspiré d'autres instruments règlementaires du marché unique numérique. Le troisième et dernier corpus prévoit des mesures en faveur de l'innovation, prenant notamment la forme de bacs à sable règlementaires[33] et d'un accompagnement/allègement règlementaire pour les PME et les *start-ups*[34].

Ensuite, plusieurs exclusions matérielles sont prévues par le texte. Il en va ainsi, premièrement, des systèmes d'IA à haut risque qui s'inscrivent dans les huit législations d'harmonisation de l'annexe I, section B[35]. L'AI Act se limite à amender ces textes sectoriels, afin qu'ils prévoient la prise en compte des exigences essentielles de l'IA de confiance[36], lorsque les produits concernés intègreront un système d'IA à haut risque, le plus souvent en tant que composant de sécurité. Deuxièmement, l'AI Act s'est clairement aligné sur la position défendue par le Conseil de l'Union visant à renforcer l'exclusion matérielle en matière de sécurité nationale et de défense, y compris militaire, afin de préserver tant l'autonomie stratégique de l'Union que la compétence des États membres en matière de sécurité intérieure[37]. Cette exclusion est certainement justifiée par la nécessité d'adopter une future règlementation des systèmes d'armes létales autonomes (SALA) – recourant à l'IA – à l'échelle internationale (et non régionale). Elle ne signifie pas pour autant que l'AI Act n'aura aucun impact en matière de défense[38].

---

[27] Selon l'article 51, §2, lorsque le « *volume cumulé de calcul utilisé pour [l']apprentissage [du modèle], mesuré en opérations à virgule flottante [FLOP], est supérieur à $10^{25}$* », comme c'est le cas des plus grands *chabots* d'IA générative (*ChatGPT-4* ou *Gemini* par ex.) aujourd'hui.
[28] Art. 51, §1, b) et 52, §1 *in fine*, AI Act.
[29] Art. 51, §2, AI Act.
[30] V. cons. 9, AI Act. Cf. proposition de directive sur la responsabilité en matière d'IA, COM/2022/496 final.
[31] Art. 1er, § 2, AI Act.
[32] *Ibid.*
[33] Art. 57 à 61, AI Act.
[34] Art. 62 et 63, AI Act.
[35] Art. 2, §2, AI Act.
[36] V. la liste de ces exigences à la section 2 du chap. III de l'AI Act, par renvoi de l'art. 102 à 110, AI Act.

Troisièmement, l'AI Act préserve la recherche et développement (R&D), y compris les sorties des systèmes ou modèles d'IA développés et mis en service uniquement à cette fin, au nom de l'innovation et de la liberté scientifique[39]. Quant aux systèmes d'IA publiés sous licences libres, ils sont également exclus de l'AI Act[40].

**2) Les frontières spatiales de l'AI Act**

Dans la droite ligne des textes NLF qui appréhendent largement la chaîne d'approvisionnement transnationale des produits mis sur le marché européen, sous l'angle tant de la responsabilité des acteurs que de la surveillance de marché[41], l'AI Act a une portée extra-européenne assumée. Les opérateurs technologiques les plus puissants sont originaires d'États tiers à l'Union. Aussi, l'AI Act se doit d'assurer un *level playing field* au sein de l'industrie de l'IA et de protéger les droits et libertés des citoyens de l'Union. Partant, le règlement s'applique, d'une part, aux fournisseurs et déployeurs établis ou situés dans l'Union et, d'autre part, aux fournisseurs établis ou situés hors de l'Union, lorsqu'ils mettent sur le marché européen, y compris à titre gratuit[42], un système d'IA ou un modèle d'IAUG couvert par l'AI Act[43]. Les fournisseurs extra-européens ont, par ailleurs, l'obligation de nommer un mandataire installé dans l'Union[44].

De manière plus originale, l'AI Act prévoit également que lorsque tant le fournisseur que le déployeur sont établis ou situés dans un pays tiers, mais que « *les sorties produites par le système d'IA sont utilisées dans l'Union* », le cadre règlementaire est applicable[45]. Sont visés ici les systèmes d'IA de nature numérique développés et déployés hors de l'Union, notamment dans un schéma de sous-traitance organisé par un opérateur européen auprès d'un opérateur établi dans pays tiers[46]. Les sorties de ces systèmes étant utilisées dans le marché européen, l'AI Act devrait s'appliquer dans ces chaînes de valeur « euro-international » de l'IA et, en amont, aux fournisseurs et déployeurs non européens. La double difficulté ici est celle de la prévisibilité et de la praticabilité de cette délimitation, spécialement pour les fournisseurs, à moins d'un alignement (futur) plus général des opérateurs mondiaux sur le référentiel de l'AI Act.

---

[37] V. art. 4, §2 TUE.
[38] En ce sens, F. Santopinto, « L'UE, l'intelligence artificielle militaire et les armes létales autonomes », *Note, Institut de relations internationales et stratégiques*, avril 2024. D'une part, l'AI Act inclut les systèmes d'IA à usage mixte – civil et militaire –, intégrés dans des drones par exemple (cons. 24). D'autre part, la dimension « humaniste » de l'AI Act qui ancre le développement et le déploiement des systèmes d'IA dans le respect des valeurs de l'Union pourrait avoir une influence sur l'approche européenne des SALA, y compris dans le cadre de sa nouvelle stratégie industrielle de défense.
[39] Art. 2, §6 de l'AI Act.
[40] Art. 2, §12, AI Act.
[41] §1.2, Guide Bleu préc.
[42] Sur les notions NLF de « *mise sur le marché* » et de « *mise à disposition* », v. art. 3, pts 9 et 10, AI Act.
[43] Art. 2, AI Act. Sont en revanche exclus les fournisseurs ou déployeurs extra-européens qui sont des personnes publiques (autorités publiques de pays tiers ou organisations internationales) utilisant des systèmes d'IA dans le cadre de la coopération internationale avec l'Union notamment en matière de services répressifs et judiciaires, dans la limite d'une protection adéquate des droits fondamentaux et libertés des personnes (art. 2, §4, AI Act).
[44] Art. 22, AI Act.
[45] Art. 2, §1, sous c), AI Act (à l'exclusion de la finalité militaire, en cohérence avec l'exclusion générique de ce domaine du cadre juridique de l'AI Act, v. art 2, §3, al. 3, AI Act).

**II) Les régimes de conformité des systèmes et modèles d'IA**

Le régime de conformité des systèmes d'IA à haut risque s'inscrit dans le sillage de la sécurité des produits (A). Par contraste, les autres systèmes de la taxonomie, y compris les modèles d'IAUG (B) ont des régimes très parcellaires[47], créant un sentiment de hiatus règlementaire.

**A) Le régime de conformité des systèmes d'IA à haut risque**

**1) Les exigences essentielles**

La conformité des systèmes d'IA à haut risque prend appui sur le concept NLF d'exigences essentielles qui sont codifiées dans la section 2 du chapitre III. Elles constituent le cœur du règlement et certainement sa partie la plus connue. Elles sont très largement inspirées des caractéristiques de l'IA de confiance dégagées, au titre de l'éthique de l'IA par le *High Level Expert Group on AI*[48], aux fins d'être transcrites ici en droit contraignant. Il s'agit principalement des exigences de gouvernance des données, de traçabilité, de transparence, de contrôle humain et des caractéristiques techniques d'exactitude, de robustesse et de cybersécurité[49]. Elles sont complétées par la dimension règlementaire de la conformité *ex ante* assurée par la mise en place, par le fournisseur, d'un système de gestion des risques[50], d'une documentation technique[51] et d'un système de gestion de la qualité[52], en vue de l'évaluation de conformité et du marquage CE.

L'ensemble de ces exigences essentielles a vocation à être traduit dans des normes techniques sous le format européen des « *normes harmonisées* »[53]. Une question délicate se pose quant aux intérêts publics que ces futures normes ont vocation à assurer. De manière originale l'AI Act vise de manière récurrente, notamment en matière de gestion des risques, la protection des droits fondamentaux. Or, comment les normes techniques peuvent-elles assurer le respect des droits fondamentaux et, plus largement, est-ce leur rôle ?[54] A l'aune de la *ratio legis* de l'AI Act qui s'émancipe de la logique « sécurité des produits » pour y ajouter une dimension « humaniste » fondée sur le concept « *[d'IA] axée sur l'humain et digne de confiance* »[55], une prise en considération, par les normes techniques, des interférences que les systèmes d'IA à haut risque ont avec les droits fondamentaux des citoyens, nous paraît nécessaire[56]. Plus fondamentalement encore, l'articulation entre le corpus européen de normalisation de l'IA[57] et celui en développement au niveau international doit retenir la plus grande attention[58]. Si un dialogue et une approche internationale interopérable, voire harmonisée, de certaines normes fondatrices (à l'instar de la terminologie, de la gestion de la qualité ou du système de management) est certainement souhaitable, cela ne doit pas conduire à un abaissement du niveau de protection prévu par l'AI Act.

---

[46] Cons. 22, AI Act.
[47] S'agissant des pratiques interdites, leur régime est celui, en principe, de l'interdiction de commercialisation, même celle-ci n'est jamais « pure et simple » (v. *supra*) et non exempt de la surveillance de marché. En outre, une disposition éparse relative à la transparence (art. 50, AI Act) s'applique de manière transverse à certains cas d'usage sans mention d'une catégorie de système d'IA à risque modéré.
[48] V. lignes directrices en matière d'éthique pour une IA digne de confiance rédigées par le groupe d'experts de haut niveau sur l'intelligence artificielle, 8 avril 2019.
[49] Art. 12 à 15, AI Act.
[50] Art. 9, AI Act.
[51] Art. 11 et annexe IV, AI Act.
[52] Art. 17 et annexe VII, AI Act.
[53] Art. 40, AI Act (et définies à l'art. 2, §1, pt c), du règlement (UE) n° 1025/2012 relatif à la normalisation). Cf. M. Ho-Dac, « La normalisation, clé de voûte de la réglementation européenne de l'intelligence artificielle (*AI Act*) », *Dalloz IP/IT*, 2023, p. 228.

**2) Les obligations des opérateurs**

L'AI Act fixe des obligations pour l'ensemble des opérateurs dans la chaîne de valeur d'un système d'IA à haut risque[59]. Sont ainsi prévues, à côté des obligations du fournisseur et du déployeur, des obligations pour les mandataires des fournisseurs[60], pour les importateurs[61] et pour les distributeurs[62]. L'éclatement transnational des opérateurs intervenant dans le développement et le déploiement d'un système ne doit pas conduire à une dilution des responsabilités et à un affaiblissement subséquent du cadre règlementaire[63].
S'agissant du fournisseur qui développe et commercialise le système, il est soumis à diverses obligations (classiques, à l'aune du NLF) visant à assurer la mise en conformité du système d'IA tout au long de son cycle de vie. La focale principale est celle de l'atténuation des risques que le système pourrait générer[64], complétée par la nécessité de coopération avec les autorités de surveillance[65]. L'AI Act instaure, par ailleurs, une obligation d'enregistrement des systèmes à haut risque par les fournisseurs sur une future base de données européenne[66]. Cette obligation doit être saluée en ce qu'elle favorisera la transparence et l'on peut regretter les dérogations importantes en matière sécuritaire[67]. Enfin, l'AI Act prévoit deux modules d'évaluation de conformité : soit l'évaluation de conformité fondée sur le contrôle interne[68], soit une procédure fondée sur la gestion de la qualité[69]. En pratique, il ouvre largement la voie au contrôle interne par les fournisseurs qui respecteront les normes harmonisées. Il place donc une grande confiance tant dans les industriels de l'IA que dans la normalisation technique. L'intervention des organismes notifiés aurait, tout au moins, pu être imposée pour les systèmes d'IA à haut risque de l'annexe III déployés à grande échelle[70]. Une autre fragilité ressort des dérogations admises à la procédure d'évaluation de conformité notamment pour des motifs de « *sécurité publique* »[71], y compris en matière répressive.

---

[54] Sur ce débat et la critique de l'approche « produits » fondée sur la normalisation en matière d'IA, v. not. M. Ebers, « Standardizing AI – The Case of the European Commission's Proposal for an Artificial Intelligence Act », L. *DiMatteo, C. Poncibò & M. Cannarsa (Eds.), The Cambridge Handbook of Artificial Intelligence: Global Perspectives on Law and Ethics,* Cambridge University Press, 2022; N. A. Smuha & K. Yeung, "The European Union's AI Act: beyond motherhood and apple pie?", *op. cit.,* p. 21 et s.
[55] Art. 1, §1, AI Act.
[56] Sans que cela ne préjuge pas des faiblesses institutionnelles qui pèsent sur le processus de normalisation. V. M. Ho-Dac, « Considering Fundamental Rights in the European Standardisation of Artificial Intelligence: Nonsense or Strategic Alliance? », in K. Jacobs (Ed.), *Joint Proceedings EURAS & SIIT 2023*, Verlag Günter Mainz, 2023. Cf. CJUE, 5 mars 2024, C-588/21 P, *Public.Resource.Org Inc*, ECLI:EU:C:2024:201 (l'existence d'un intérêt public supérieur justifie l'accès gratuit à des normes harmonisées sur la sécurité des jouets en ce qu'elles font partie intégrante du droit de l'Union).
[57] Ces travaux sont actuellement conduits au sein du CEN-CENELEC *JTC* 21.
[58] V. spéc. les travaux conduits au sein de ISO/IEC *JTC* 1/*SC* 42.
[59] V. chap. III, section 3, art. 16 à 27, AI Act.
[60] Art. 22, AI Act.
[61] Art. 23, AI Act.
[62] Art. 24, AI Act.
[63] Cons. 83, AI Act. Aussi, selon l'article 25 de l'AI Act, un distributeur, importateur, fabricant de produit ou déployeur peut être assimilé, à certaines conditions, à un fournisseur de système d'IA et soumis aux obligations de ce dernier au titre de l'article 16.
[64] V. spéc. art. 20, AI Act.
[65] Art. 21, AI Act.

Quant au déployeur, on relèvera d'emblée qu'il est un destinataire essentiel du cadre règlementaire de l'AI Act, aux côtés du fournisseur ; toutes les entités – privées comme publiques – qui recourent à l'IA doivent en avoir bien conscience. Ainsi, les déployeurs sont associés à la mise en œuvre (et donc au respect) de plusieurs exigences essentielles qui fondent le cadre de confiance des systèmes d'IA à haut risque : en matière de contrôle humain[72], de qualité des données[73] ou encore de traçabilité[74] et de transparence sous la forme d'un devoir d'information des personnes concernées lorsque le déploiement du système à haut risque est utilisé pour une (aide à la) prise de décision[75] ou sur le lieu de travail[76]. En creux, il s'agit de protéger les citoyens contre les risques de discrimination, d'atteintes à l'autodétermination informationnelle ou à la vie. Dans le même esprit, certains déployeurs devront conduire une étude d'impact sur les droits fondamentaux pour certains systèmes d'IA[77]. Il faut se féliciter de cette disposition, insérée sous la pression du Parlement européen, même si son champ d'action personnel a été restreint aux seules administrations publiques, à l'exception de certains cas d'usage particulièrement sensibles dans le domaine des services financiers. Ces mêmes déployeurs « personnes publiques » sont également astreints à un obligation d'enregistrement en sélectionnant, sur la base de données européenne précitée, les systèmes d'IA à haut risque – enregistrés préalablement par le fournisseur – qu'ils vont utiliser[78]. Enfin, les déployeurs ont un rôle important sur le terrain du *public enforcement*, puisqu'ils sont associés au monitoring de la sécurité des systèmes qu'ils déploient, en dialogue avec les fournisseurs et les autorités de surveillance[79]. Le déployeur est, en outre, responsable de la bonne exécution du droit à l'explication dont les utilisateurs finaux sont titulaires, à certaines conditions plutôt resserrées, en présence d'une décision prise par ce dernier sur la base d'un système à haut risque[80].

---

[66] Art. 49, 71 et annexe VIII, AI Act.
[67] Art. 49 §4, AI Act.
[68] Annexe VI, AI Act.
[69] Annexe VII, AI Act.
[70] Il est d'ailleurs dommage et même peu compréhensible (si ce n'est pour des raisons politiques) que le concept de « *risque systémique* » soit cantonné aux modèles d'IAUG qui, paradoxalement, ont un régime allégé par rapport aux systèmes d'IA à haut risque.
[71] Art. 46, AI Act. Dans cette dernière hypothèse, le système d'IA pourrait être utilisé par les autorités répressives en amont de toute évaluation de conformité et sans autorisation préalable de l'autorité nationale de surveillance : cela ne paraît pas raisonnable à l'aune du respect des valeurs de l'Union.
[72] Art. 14, AI Act.
[73] Art. 26, §4, AI Act.
[74] Art. 26, §6, AI Act.
[75] Art. 26, §11, AI Act.
[76] Art. 26, §7, AI Act.
[77] Art. 27, AI Act.
[78] Art. 26, §8 et 49, §3, AI Act.

**B) Le régime de conformité des modèles d'IAUG**

Les modèles d'IAUG font l'objet, en cas de mise sur le marché, d'un régime de conformité à plusieurs niveaux selon, d'une part, qu'ils sont ou non diffusés sous licence libre et, d'autre part, qu'ils présentent ou non des risques systémiques. Quel que soit son niveau, ce régime vient en principe s'ajouter au régime de conformité applicable au système d'IA auquel le modèle est intégré, que le fournisseur de ce système et de ce modèle soit, ou non, le même opérateur[81] ; l'enjeu de ce cumul règlementaire est central au stade de la mise sur le marché d'un système d'IA à haut risque – qu'il soit ou non à usage général – ayant intégré un tel modèle, afin d'assurer la protection des droits des citoyens et de la société. A ce titre et aux fins d'assurer l'effectivité extra-européenne de cette protection, les fournisseurs non européens de modèles d'IAUG doivent, eux aussi, désigner un mandataire établi dans l'Union[82].

S'agissant du premier niveau de conformité qui vise tous les modèles d'IAUG, sauf ceux sous licence libre qui en sont exemptés[83], il prend la forme d'un régime de transparence principalement fondé sur l'élaboration d'une documentation technique, à destination tant des autorités de régulation – en l'occurrence le « *Bureau de l'IA* » – que des fournisseurs de systèmes d'IA qui envisagent d'intégrer un tel modèle. L'annexe XI de l'AI Act fournit un canevas pour réaliser la documentation technique du modèle et l'annexe XII déroule le contenu minimum de l'obligation d'information dû aux fournisseurs de systèmes d'IA. Ces obligations devraient être opérationnalisées, dans leur dimension technique, par de futures normes harmonisées[84], complétées par des actes délégués de la Commission[85]. Les fournisseurs de modèles d'IAUG doivent également informer le public « *du contenu utilisé pour entraîner le modèle* » sur la base d'un formulaire-type que le Bureau de l'IA devra élaborer. Enfin, ils sont invités à mettre en place une politique de conformité à l'aune des exigences européennes du droit d'auteur.

Quant au second niveau de conformité, il vise les fournisseurs de modèles d'IAUG présentant un risque systémique. Ils ont une obligation de notification de leur modèle auprès de la Commission européenne qui devra assurer la publication d'une liste de ces modèles, dans le respect du secret des affaires[86]. Ils doivent, en outre, mettre en place des politiques de gestion des risques, en écho à l'exigence de l'article 9 en matière de systèmes à haut risque. Il s'agit d'évaluer les modèles suivant une méthodologie d'étude d'impact permettant d'identifier les risques systémiques et, le cas échéant, de les atténuer. Sur le modèle NLF de l'article 73 de l'AI Act, les fournisseurs de modèles présentant des risques systémiques ont également une obligation de suivi et de communication au Bureau de l'IA de tout incident grave et des mesures correctives prises en réponse[87].

---

[79] Art. 26, §5, AI Act. Les déployeurs sont chargés d'informer le fournisseur et les autorités de surveillance en cas de système présentant un risque pour les intérêts publics protégés par l'AI Act (art. 79, §1) et en cas d'incident grave (art. 73).
[80] Art. 86, AI Act.
[81] Cons. 97, AI Act.
[82] Art. 54, AI Act.
[83] Art. 53, §2 et cons. 104, AI Act.
[84] Des codes de bonne pratique devraient être préparés sous la supervision du Bureau de l'IA dans l'attente de l'élaboration desdites normes techniques (art. 56, AI Act)
[85] Art. 53, §4 et 5, AI Act.
[86] Art. 52, §6, AI Act. Sur cette obligation, v. *supra*, I) A) 3).

**III ) La gouvernance multi-niveaux des systèmes d'IA**

Le cadre de *public enforcement* de l'AI Act est holistique. Le texte prend appui sur le schéma de surveillance du marché au sens du règlement (UE) 2019/1020 qui s'applique « *dans son intégralité* »[88] mais va au-delà en établissant une gouvernance multi-niveaux. Cela est justifié par la volonté « *de renforcer les capacités au niveau de l'Union et d'intégrer les parties prenantes dans le domaine de l'IA* »[89]. Il y a ainsi une volonté centralisatrice à l'échelle fédérale européenne, dans la lignée d'autres grands textes et domaines stratégiques du marché intérieur. En appui de ce niveau d'exécution européen (B), le niveau national demeure central (A).

**A) La gouvernance nationale décentralisée fondée sur la surveillance de marché**

Pour ce qui, d'abord, de la dimension institutionnelle de la surveillance de marché, l'AI Act laisse aux États membres une marge de manœuvre dans la désignation des autorités nationales compétentes. Chaque État doit se doter, « *au moins* », d'une autorité notifiante et d'une autorité de surveillance du marché, avec comme principes-clés d'exercice de leur compétence, l'indépendance et l'impartialité[90]. D'un côté, les autorités notifiantes ont pour mission d'évaluer, de désigner, de notifier et de contrôler les organismes en charge de l'évaluation de la conformité des systèmes d'IA à haut risque[91]. Ces organismes (précités) sont au cœur du schéma de conformité *ex ante* des fournisseurs de systèmes d'IA à haut risque puisque ce sont eux qui doivent superviser le module de conformité fondé sur l'évaluation du système de gestion de la qualité[92]. Leur régime est très détaillé en vue de garantir leur haut niveau d'expertise et d'indépendance[93]. De l'autre côté, les autorités nationales de surveillance jouent le rôle de régulateur du marché domestique au sein duquel les systèmes d'IA sont mis à disposition ; cette mission pourra être partagée entre plusieurs entités nationales, selon les « *besoins organisationnels* » des États membres. Dans tous les cas, un « *point de contact unique* » devra être désigné, dans chaque État membre, afin de jouer le rôle d'interface avec le public et dans les relations avec l'Union et les autres États membres[94]. En outre, l'AI Act impose (avec des dérogations possibles), pour certains domaines à haut risque, la désignation d'autorités de contrôle préexistantes ; c'est le cas, d'une part, dans le cadre des législations sectorielles NLF (annexe I, section A) et en matière de services financiers afin d'éviter un éclatement régulatoire[95] et, d'autre part, s'agissant de cas d'usage de l'annexe III hautement sensibles pour les droits fondamentaux (matière répressive, gestion des frontières et justice, y compris dans le contexte d'un système biométrique) qui devraient relever des autorités de protection des données. Finalement, s'agissant des systèmes d'IA développés ou déployés par des institutions ou organes de l'Union, c'est le Contrôleur européen de la protection des données (CEPD) qui agira en tant qu'autorité de surveillance[96].

---

[87] Art. 55, §1, sous c), AI Act.
[88] Cons. 156 et art. 74, AI Act.
[89] Cons. 148, AI Act.
[90] Art. 70, AI Act.
[91] En France, la COFAC est l'autorité notifiante de référence en matière de sécurité des produits.
[92] Art. 43 et annexe VII, AI Act.
[93] V. section IV, chap. III, AI Act. Cf. spéc. art. 36 et 37, AI Act.
[94] Ce rôle doit être assumée par une autorité de surveillance selon l'article 70, §2. En France, le Pôle normalisation et règlementation des produits (PNRP) de la Direction générale des entreprises (DGE) endosse actuellement les responsabilités de *Bureau de liaison unique* (BLU) prévu à l'article 10 du règlement « surveillance de marché ».

Pour ce qui est, ensuite, du contenu de la « *surveillance du marché* », l'AI Act renvoie au règlement éponyme et en reprend les principaux mécanismes. Les autorités nationales surveillent leur marché des systèmes d'IA à haut risque, à l'exclusion des modèles d'IA à usage général qui relèvent de la surveillance de niveau européen[97] – à l'instar des très grandes plateformes dans le cadre du *Digital Services Act* –. Elles conduisent des opérations de contrôle et, le cas échéant, des enquêtes, y compris conjointes avec la Commission européenne[98]. Elles sollicitent la prise de mesures correctives par les fournisseurs en cas de non-conformité règlementaire et prennent des mesures en cas de refus de mise en conformité par ces derniers[99]. C'est également le cas pour les systèmes « *présentant des risques* », y compris en cas de système conforme à l'AI Act[100]. La Commission est tenue informée afin qu'une coordination entre autorités nationales puisse être mise en œuvre. Dans le cadre de l'exercice de leurs missions, les autorités nationales pourront obtenir des fournisseurs un accès complet à la documentation, y compris aux jeux de données d'entraînement, de validation et de test et, à certaines conditions, au code source du système[101]. L'AI Act prévoit, par ailleurs, un mécanisme de réclamation assez largement ouvert auprès de l'autorité de surveillance du marché « *concernée* », en cas de violation des dispositions du règlement[102].

Quant au régime de sanctions des opérateurs en cas de violation des obligations prévues par l'AI Act, il est assez précis et donc uniformément dissuasif : des amendes de 15 millions d'euro ou, pour les entreprises, de 3% de leur chiffre d'affaires annuel mondial peuvent être imposées aux opérateurs[103] et ces montants atteignent 35 millions d'euro et 7% de leur chiffre d'affaires en cas de violation des pratiques d'IA interdites[104].

**B) Le cadre européen de gouvernance de l'IA**

---

[95] Art. 74, §3 et §6, AI Act.
[96] Art. 74, §8, AI Act.
[97] Des règles de coordination de la supervision entre les autorités nationales et le Bureau de l'IA sont prévues en cas d'interrelation entre un système et un modèle d'IA à usage général, v. art. 75 de l'AI Act. Lorsque le système et le modèle d'IAUG sont développés par un même fournisseur, la supervision est confiée au Bureau de l'IA (§1, art. 75). La supervision d'un système d'IAUG relève des autorités nationales en coopération avec le Bureau de l'IA (§2, art. 75). Les autorités nationales peuvent également obtenir de l'aide du Bureau de l'IA pour conduire des enquêtes relatives à un système prenant appui sur un modèle d'IAUG.
[98] Art. 74, AI Act. Cf. art. 9, 11, 14 à 21, règl. « surveillance de marché ».
[99] Cf. art. 11, règl. « surveillance de marché ».
[100] Art. 79 et 82, AI Act.
[101] Art. 74, §§12 et 13, AI Act.
[102] Art. 85 de l'AI Act.
[103] Art 99, §4, AI Act.
[104] Art. 99, §3, AI Act.

En surplomb de la surveillance nationale de marché, l'AI Act instaure une gouvernance européenne de l'IA composée de trois entités : une entité « verticale », au sein de la Commission européenne et dénommé « *Bureau de l'IA* »[105] ; une entité « horizontale » sous le nom de « *Comité européen de l'IA* »[106], composée des représentants des régulateurs des États membres, du Bureau de l'IA (sans droit de vote) et de membres soit observateurs – à l'instar du CEPD – soit invités ; et enfin, une entité « experte » composée d'un « *Forum consultatif* »[107] réunissant des représentants des parties prenantes de l'écosystème de l'IA en vue de conseiller le Comité européen de l'IA et la Commission, et d'un « *Groupe scientifique* »[108] dont la mission est de soutenir la mise en œuvre de l'AI Act, grâce à des connaissances scientifiques de pointe, notamment en matière de modèles d'IAUG.

S'agissant du Bureau de l'IA, il est une émanation de la Commission avec une double fonction : d'une part, la surveillance des modèles d'IA à usage général et, d'autre part, la gouvernance européenne de l'IA dans le contexte transnationale du marché intérieur de l'Union[109]. Sa fonction de superviseur des modèles d'IA est largement calquée sur celle des autorités nationales de surveillance à l'égard des systèmes d'IA à haut risque : adopter des mesures de contrôle et conduire des enquêtes, donner suite aux alertes de risques systémiques en conduisant une évaluation de conformité du modèle, accéder à la documentation technique, solliciter des fournisseurs des mesures correctives notamment en cas de non-conformité[110], mettre en place un système de réclamations pour les fournisseurs en aval[111] et prononcer des sanctions sous forme d'amendes[112]. De manière plus spécifique, le Bureau de l'IA est en charge d'établir la méthodologie de classification des modèles d'IA à usage général[113] et de l'élaboration des codes de bonne pratique qui doivent jouer comme base-relais de la mise en conformité dans l'attente des futures normes techniques en matière d'obligation de transparence des modèles d'IA[114]. Quant à sa mission de gouvernance de haut niveau, elle est éclatée à travers diverses dispositions de l'AI Act. Elle consiste, d'une part, à élaborer différents documents de référence favorisant une mise en œuvre harmonisée et cohérente du texte par les parties prenantes[115]. D'autre part, le Bureau de l'IA se voit confier un rôle de coordinateur des différents acteurs – industriels comme institutionnels – et de promotion de la coopération transfrontière[116], y compris internationale. Ce dernier aspect est central dans le contexte d'institutionnalisation progressive d'une gouvernance mondiale de l'IA notamment dans sa dimension « sécurité » (*safety*). Dans l'ensemble de ces tâches, le Bureau devrait recevoir le support technique du *Groupe scientifique* précité.

---

[105] Art. 64, AI Act. V. décision de la Commission du 24 janvier 2024, créant le bureau de l'intelligence artificielle, C/2024/1459.
[106] Art. 65, AI Act.
[107] Art. 67, AI Act.
[108] Art. 68, AI Act.
[109] Art. 3, pt 47, AI Act.
[110] Art. 89 à 93, AI Act.
[111] Art. 89, §2 (pour les fournisseurs en aval) et cons. 162, AI Act.
[112] Art. 101, §1, AI Act qui fixe le montant à 3 % de du chiffre d'affaires annuel mondial ou 15 millions d'euro à l'encontre des fournisseurs de modèles d'IAUG.
[113] Art. 55, §1 sous c), AI Act.
[114] Art. 56 et cons. 116, AI Act.
[115] Par ex. un questionnaire pour l'étude d'impact sur les droits fondamentaux que certains déployeurs devront réaliser (art. 27, §5) et un modèle de résumé du contenu utilisé pour entraîner les modèles d'IAUG (art. 53, §1, sous d).
[116] Par ex. art. 74, §11, AI Act.

S'agissant du Comité européen de l'IA, il a un rôle d'appui de la Commission et des États membres qu'il « *conseille et assiste* »[117] sur une multitude de sujets qui sont fédérés par l'objectif de favoriser la bonne mise en œuvre du cadre règlementaire. En ce sens, le Comité européen peut intervenir pour accompagner l'application cohérente de l'AI Act en soutenant la coordination des activités des autorités nationales, y compris l'échange de bonnes pratiques et l'harmonisation des procédures administratives sur certaines questions (dérogations à la procédure d'évaluation de conformité, bacs à sable règlementaires). Il a également une vocation plus politique qui ne devrait pas, on l'espère, concurrencer le Bureau de l'IA – en tant que voix de l'Union dans le concert mondial de la gouvernance de l'IA –, s'agissant du développement de la coopération internationale, de prises de position sur la géopolitique mondiale de l'IA ou encore sur le régime applicable aux modèles d'IAUG.

\* \* \*

L'enjeu central de la mise en œuvre effective de l'AI Act est à présent entre les mains des différentes parties prenantes ; elle nécessitera une lecture cohérente et équilibrée de ses dispositions qui sont aux prises avec des intérêts antagonistes parfois difficiles à concilier : innovation et liberté d'entreprendre, garantie des droits et des libertés, autonomie stratégique et sécurité intérieure. Dans ce contexte, l'Union devra se montrer à la hauteur de son *leadership* règlementaire guidé par ses valeurs fondatrices.

---

[117] Art. 66, AI Act.